# Superconductivity in the $Mn_5Si_3$-type $Zr_5Sb_3$ system

B. Lv,[1]* X. Y. Zhu,[1] B. Lorenz,[1] F. Y. Wei,[1] Y. Y. Xue,[1] Z. P. Yin,[2] G. Kotliar[2] and C. W. Chu[1,3]*

[1]Department of Physics and Texas Center for Superconductivity, University of Houston, Houston TX 77204-5002

[2]Department of Physics and Astronomy, Rutgers University, Piscataway, NJ 08854-8019

[3]Lawrence Berkeley National Laboratory, Berkeley CA 94720

* To whom correspondence should be addressed. E-mail: blv@uh.edu or cwchu@uh.edu.

Abstract:

We report the discovery of superconductivity at 2.3 K in $Zr_5Sb_3$, the first superconducting member in the large compound family of the $Mn_5Si_3$-structure type. Transport, magnetic, and calorimetric measurements and band structure calculations show it to be a phonon-mediated BCS superconductor, with a relatively large density of states at the Fermi level associated with the d-electrons of Zr and substantially larger electron-phonon coupling compared to the Sn counterpart compound $Zr_5Sn_3$. More superconductors with even higher transition temperatures are expected to be found in this family of compounds.

Searching for superconductivity in materials of new distinct structural classes has proven to be fruitful in achieving superconductivity with higher transition temperature $T_c$ or discovering novel mechanisms. For instance, it was the unexpected observation of superconductivity in 1979 in $CeCu_2Si_2$ that heralded in the exciting systems of heavy-fermions, in spite of its initial low $T_c \sim 1$ K.[1] It was the discovery of superconductivity in the omnipresent perovskite-like layered cuprates, first at 35 K in Ba-doped $La_2CuO_4$ in 1986[2] and then at 93K in $YBa_2Cu_3O_7$ in 1987,[3] that inaugurated the modern era of high temperature superconductivity for exciting science and technology in the ensuing decades till today. Then it was the detection of superconductivity in 2008 in layered Fe-based compounds, F-doped LaFeAsO with an unusually high $T_c \sim 26$ K,[4] that has opened up a new opportunity in unraveling the high temperature superconducting mechanism and has offered new hopes for superconductors of high $T_c$ due to the presence of a large amount of magnetic Fe-element and a large number of compounds in this structural class. We have therefore decided to explore compounds of a structural class that consists of large number of compounds where superconductivity is likely but has not yet been found.

We have chosen to investigate the binary $A_5B_3$ system where A = early transition metal or rare earth elements and B = IIIB–VB elements. The compound family, $A_5B_3$, has been shown to consist of more than 590 compounds with three distinct structure-types, i.e. hexagonal $Mn_5Si_3$-type ($P6_3/mcm$) with ~ 440 compounds, tetragonal $Cr_5B_3$-type ($I4/mcm$) with ~ 86 compounds, and $Yb_5Sb_3$-type (Pnma) with ~ 65 compounds.[5] There exists no report of superconductivity in the large compound family of $Mn_5Si_3$ type to date. In view of the possible large electron density of states associated with the d-electrons in the transition metal elements that may give rise to superconductivity, we have carried out detailed structural, magnetic, transport, and calorimetric studies of the hexagonal $Mn_5Si_3$ structure-type $Zr_5X_3$ compounds for X = Sb, Sn, Ge, Ga, and Al. Superconductivity at ~ 2.3 K has been discovered in $Zr_5Sb_3$, although not in the other members tested, possibly due to their much smaller electron density of states and weaker electron-phonon coupling. $Zr_5Sb_3$, therefore, represents the first superconducting member of this large compound family of $Mn_5Si_3$-type and gives hope that many others even with higher $T_c$ may be found.

Samples were prepared through arc-melting pellets of Zr and X in a nominal composition of 5:3+δ on a water-cooled copper hearth in a home-made arc furnace in argon-atmosphere with a Zr gas getter. For the volatile element X, such as Sb, δ is greater than 0, e.g. δ ~ 0.3, to compensate for the loss of Sb through vaporization when the pellet is first brought close to the arc. The X (Sb, Sn, Ge, Ga, and Al) elements are pieces from Sigma Aldrich with > 99.999% purity and are used as received. The Zr is purchased from Alfa Aesar and further purified through decomposition of $ZrH_2$ as described previously.[6,7]

X-ray powder diffraction was performed using a Panalytical X'pert Diffractometer. The dc magnetic susceptibility χ(T) and χ(H) was measured using a Quantum Design Magnetic Property Measurement System (MPMS) down to 2 K and up to 5 T. Electrical resistivity ρ(T) and ρ(H) was measured by employing a standard 4-probe method using a Linear Research LR-400 ac bridge operated at 15.9 Hz in a Quantum Design Physical Property Measurement System (PPMS) up to 7 T and down to 1.9 K. The specific heat $C_p(T)$ measurement was determined down to 0.4 K in a field up to 7 T using the Quantum Design He3-attachment in the Quantum Design PPMS.

The hexagonal $Mn_5Si_3$-type compounds have been widely studied previously in terms of their structure, stoichiometry, and host-interstitial chemistry, particularly for the Zr-based compounds.[6-9] In the hexagonal $Zr_5X_3$ (X = Sb, Ga, Ge, and Sn) compounds, there are three distinct crystallographic sites. As shown in Fig. 1, the trigonal antiprisms consisting of Zr1 atoms are face-shared and form a column along the *c* axis. The shared faces are then each bridged by three X atoms to give $Zr_{6/2}X_{6/2}$ chains. Each of the Zr1 atoms has 6 Zr and 4 X neighbor atoms. Meanwhile, a parallel linear chain comprised of Zr2 atoms is closely bonded with the $Zr_{6/2}X_{6/2}$ chain. The Zr2 atoms are neighbored with 6 X forming a twisted trigonal prism. The octahedral interstitial sites formed by the Zr1 trigonal antiprism atoms are ideal for small atoms to occupy. In fact, A variety of compounds $Zr_5Sb_3Z$ have been discovered, with interstitial guest atoms Z = C, O, Al, Si, P, S, Co, Ni, Cu, Zn, Ge, As, Se, Ru, and Ag occupying the centers of zirconium trigonal antiprismatic sites in the $Zr_5Sb_3$ host.[8] Band structural calculations have shown that the zirconium states and electrons are diverted from the broad conduction band to form the Zr-Z bonds.[8]

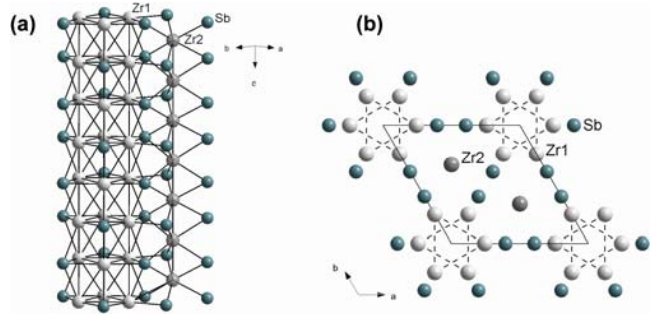

FIG. 1. Crystal structure of $Zr_5Sb_3$: (a) side view; (b) [001] projection. Two types of Zr atoms are labeled differently for distinction.

Polycrystalline samples of $Zr_5X_3$ (X = Sb, Al, Ga, Ge, and Sn) were synthesized through arc-melting method. Figs. 2(a) and 2(b) show the X-ray Rietveld refinement patterns of $Zr_5Sb_3$ and $Zr_5Sn_3$, respectively. The sharp peaks and good refinement values of Rp and Rwp (Rp = 4.60%, Rwp = 7.11% for $Zr_5Sb_3$ and Rp = 5.38%, Rwp = 7.58% for $Zr_5Sn_3$) demonstrate that they possess the hexagonal $Mn_5Si_3$ structure type. The refined lattice parameters are *a* = 8.4053(3)Å, *c* = 5.7640(3)Å for $Zr_5Sb_3$ and *a* = 8.4606(3)Å, *c* = 5.7771(2)Å for $Zr_5Sn_3$, and match well with previous studies.[7,10] A few minor peaks are also found in the X-ray powder patterns of $Zr_5Sb_3$ [marked by red stars in Fig. 2(a)], estimated to be less than 5% of the major peaks intensity and possibly due to the orthorhombic $Y_5Bi_3$-type $Zr_5Sb_3$. The metallic gray ingots after arc-melting are moderately stable for a few hours in humid air before turning into black powder.

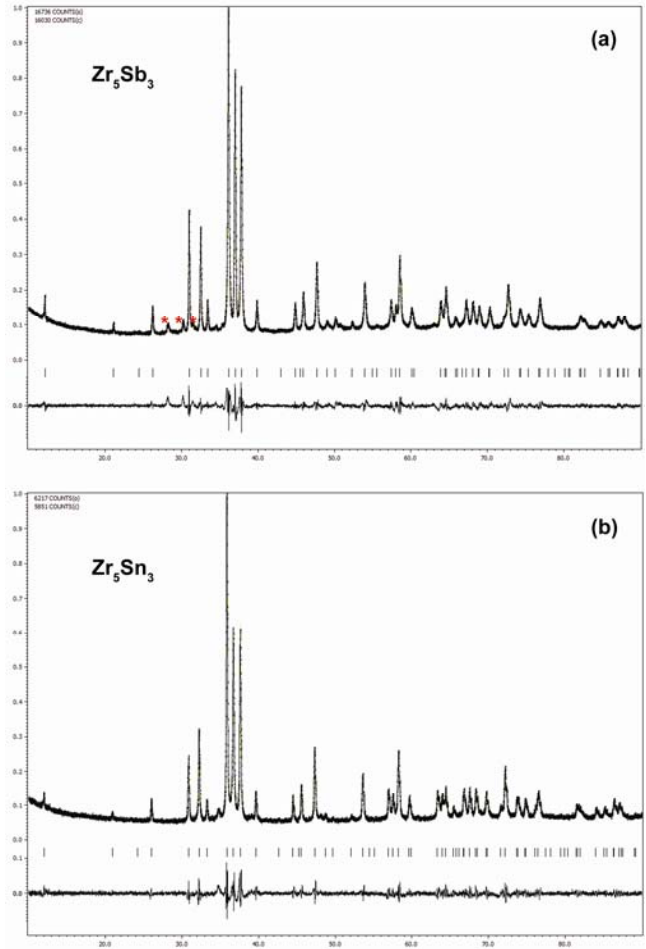

FIG. 2. Rietveld refinement of (a) $Zr_5Sb_3$ (impurity peaks marked by red stars) and (b) $Zr_5Sn_3$ x-ray powder pattern.

The superconductivity in $Zr_5Sb_3$ is evident from the magnetic susceptibility $\chi(T)$, electrical resistivity $\rho(T,H)$ and the specific heat $C_p(T)$ measurements. Figure 3 displays the $\chi(T)$ measured at 5 Oe both in the zero-field-cooled (ZFC) and field-cooled (FC) modes. A clear diamagnetic shift is observed below 2.3 K. The shielding volume fraction from ZFC-χ(T) is ~ 1.3 before demagnetization correction, in conjunction with the FC-χ(T), indicating bulk superconductivity for the sample. The M(H) curve at 1.9 K in the inset shows clear type-II superconductor characteristics, with a low critical field $H_{c1}$ < 5 Oe.

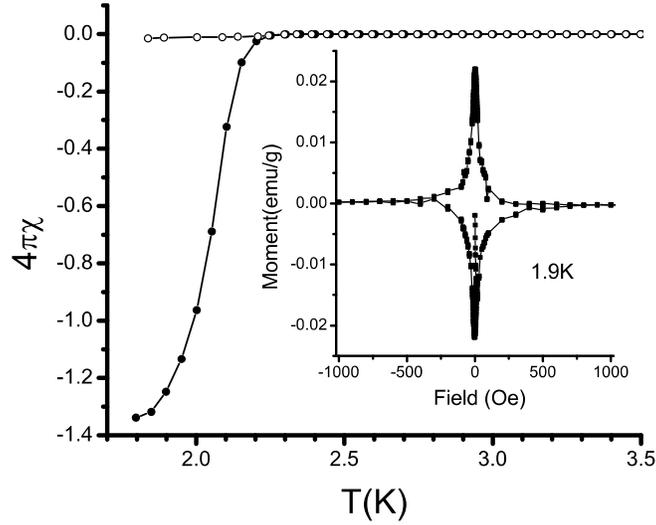

Fig. 3. Zero-field-cooled (ZFC) susceptibility (solid circles) and field-cooled (FC) susceptibility (open circles) data at 5 Oe for $Zr_5Sb_3$ sample. Inset: M-H curve at 1.9 K.

The electrical resistivity ρ(T) of $Zr_5Sb_3$ is shown in Fig. 4, with a room temperature value of 0.36 mΩ-cm. It decreases with decreasing temperature, typical for a metal, and drops sharply below 2.3 K, characteristic of a superconducting transition. The transition width is estimated to be 0.2 K from 10% to 90% resistivity drops. In the presence of magnetic fields, the superconducting transition is systematically shifted to lower temperature, and suppressed to below 1.9 K at 7 kOe, as shown in the inset of Fig. 4.

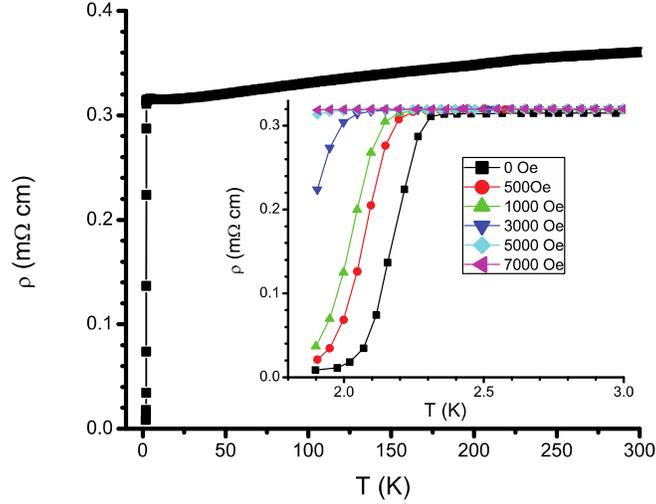

FIG. 4. Resistivity data of $Zr_5Sb_3$ at H = 0 between 1.9 K and 300 K. Inset: resistivity data under different magnetic fields between 1.9 K and 3.0 K.

The bulk superconductivity of the compound is further demonstrated by the pronounced specific heat $C_p$ anomaly, as shown in Fig. 5. The superconducting transition is suppressed to below 0.4 K in a magnetic field of 30 kOe (Fig. 5). The electronic Sommerfeld coefficient γ is 24.1 mJ/(mol K$^2$). The electronic contribution to the heat capacity, $C_{el}$, in the superconducting state is determined by subtracting the 30 kOe data. The normalized $C_{el}/(γ T)$ as a function of the reduced temperature, $T/T_c$, is plotted in the inset of Fig. 5 and compared with the theory. The BCS theory (dashed line) does not fit the data well; however, the modification of the superconducting gap as described by the α-model[11] leads to a good description of the experimental data. In magnetic fields the heat capacity peak is shifted to lower temperature, as expected for the superconducting state. Data in different external fields are shown in Fig. 6. The upper critical field, $H_{c2}(T)$, is derived from the heat capacity data and shown in the inset of Fig. 6. The extrapolation of $H_{c2}$ to zero temperature is done by fitting the data to the Werthamer-Helfand-Hohenberg (WHH) theory,[12] shown as the red line in the figure. $H_{c2}(0)$ is indeed found to be less than 30 kOe, as already suggested by the heat capacity data of Fig. 6. Unfortunately, the isostructural $Zr_5Sn_3$ was found not to be superconducting. Its $C_p$ reveals a much smaller Sommerfeld coefficient = 14 mJ/(mol K$^2$).

Since γ in normal metals is proportional to the electronic density of states (DOS) at the Fermi level $E_F$, a 40% lower value may explain the missing superconducting state in $Zr_5Sn_3$.

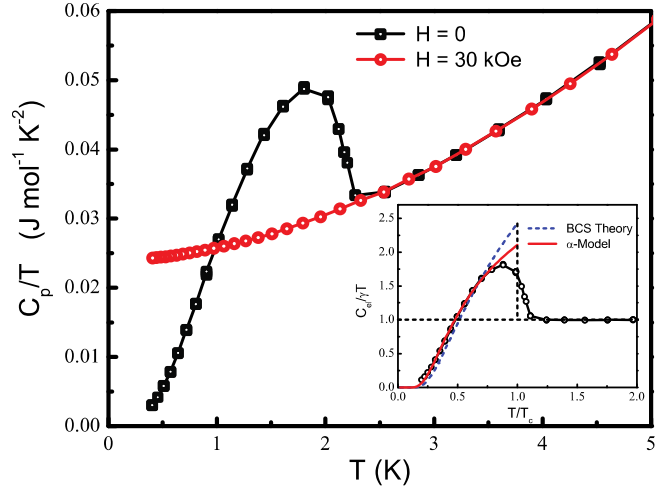

FIG. 5. Heat capacity of $Zr_5Sb_3$ at zero field (open squares) and at H = 30 kOe (open red circles). The inset shows the normalized electronic heat capacity in comparison with the BCS- and α-models (α = 1.55).

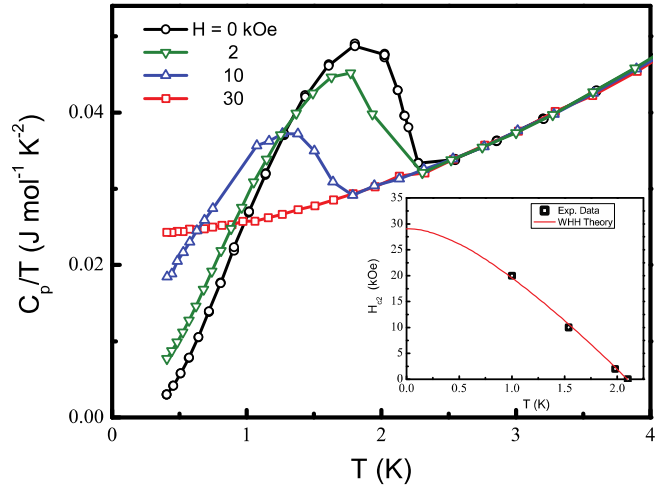

FIG. 6. Field dependence of the heat capacity of $Zr_5Sb_3$. The inset shows the upper critical field vs. temperature (open squares) and the fit to the WHH theory (red line).

The band structures and the DOS at the Fermi level of both $Zr_5Sb_3$ and $Zr_5Sn_3$ have been calculated. $Zr_5Sb_3$ has bands at the Fermi level that are much more strongly coupled to phonons than in $Zr_5Sn_3$, indicating $Zr_5Sb_3$ has a much stronger electron-phonon coupling than $Zr_5Sn_3$.[13] The DOS of $Zr_5Sb_3$ at the Fermi level is about 2.4 times higher than that of $Zr_5Sn_3$, as shown in Fig. 7, in good agreement with the results of the $C_p$-experiments. Therefore, the observed superconductivity in $Zr_5Sb_3$ is of the BCS type arising from decent electron-phonon interaction as well as its high DOS at the $F_F$ due to the d-electrons of Zr.

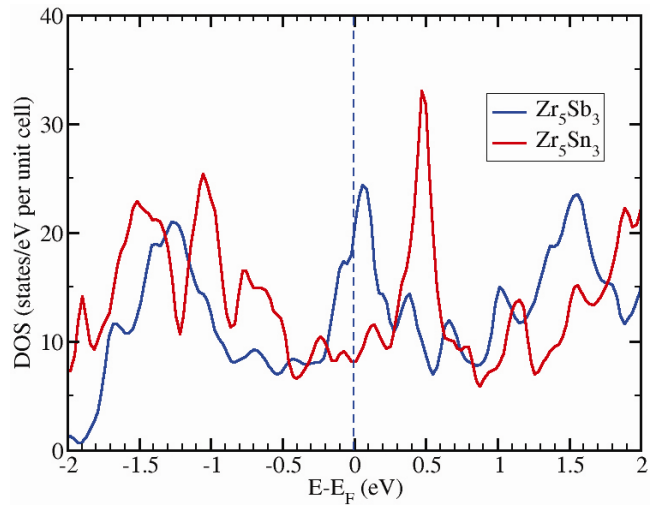

FIG. 7. The density of states (DOS) of $Zr_5Sb_3$ and $Zr_5Sn_3$ calculated by density functional theory with the generalized gradient approximation exchange-correlation functional.[14]

Both hexagonal $Mn_5Si_3$-type ($P6_3/mcm$) and orthorhombic $Y_5Bi_3$-type (Pnma) structures exist for the $Zr_5Sb_3$ compound. The former appears non-stoichiometric as $Zr_5Sb_{3+x}$ with $0 \leq x \leq 0.4$, while the latter forms stoichiometrically but only stable at high temperature above 1200 °C.[7] To further investigate how the non-stoichiometry affects the superconducting properties, samples with nominal compositions of $Zr_5Sb_3$ and $Zr_5Sb_{3.3}$ have been made through solid state sintering process. Different stoichiometric materials were mixed thoroughly and pressed into a pellet inside a glove box. The pellets were initially sealed in a Nb tube under Ar, subsequently sealed in the quartz jacket under vacuum, and then heated up very slowly to 1000 °C for 5 days. As illustrated in Fig. 8(a), both the onset $T_c$ and shielding fraction decrease in the $Zr_5Sb_{3.3}$ sample. As previous studies indicated, the extra Sb in $Zr_5Sb_{3+x}$ occupies the octahedral interstitial sites formed by the Zr trigonal antiprism,[7] and our results show that superconductivity in $Zr_5Sb_3$ is suppressed when the interstitial sites are filled by Sb atoms. We have further tested the influence of Z = O and C, which are introduced to $Zr_5Sb_3$ through arc-melting and sintering processes. Only a small diamagnetic signal below 2 K was detected in $Zr_5Sb_3O_x$ and no superconductivity was observed in $Zr_5Sb_3O$ and $Zr_5Sb_3C$ samples [Fig. 8(a)], further proving that superconductivity is not favored by interstitial-filled compounds of $Zr_5Sb_3$. This is also consistent with previous theoretical calculations that the formation of the Zr-Z bond significantly reduces the Zr1-Zr1 bonding from the trigonal antiprism construction, and thus might reduce the density of states at the Fermi surface and suppress the superconducting signal.

Chemical doping with different species has also been carried out. The superconductivity seems pretty robust and stays at ~ 2.3 K with 5% of Hf and Y doping [Fig. 8(b)]. However, 5% Ti doping has effectively suppressed the superconducting transition from 2.3 K down to 2 K [Fig. 8(b)], and there is no superconducting signal observed in the 10% Ti-doped sample.

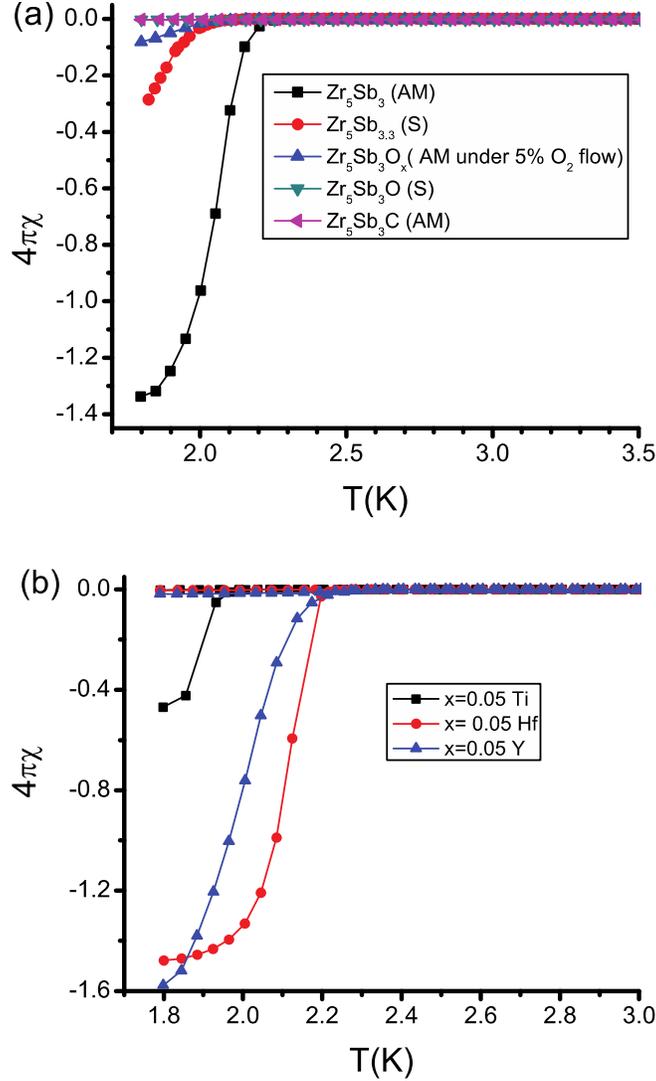

FIG. 8. (a) ZFC susceptibilities under 5 Oe field for $Zr_5Sb_3$, $Zr_5Sb_{3.3}$, $Zr_5Sb_3O_x$, $Zr_5Sb_3O$, and $Zr_5Sb_3C$. Abbreviations: AM—arc melted; S—sintered. (b) ZFC and FC susceptibilities under 5 Oe field for different $(Zr_{1-x}M_x)_5Sb_3$ (M = Ti, Hf, and Y) samples.

In summary, we have carried out magnetization, electrical resistivity, and specific heat measurements, and band structure calculations on $Zr_5Sb_3$ and $Zr_5Sn_3$. The results show bulk superconductivity with a $T_c \sim$ 2.3 K in $Zr_5Sb_3$, representing the first superconductor discovered in the large family of $Mn_5Si_3$ structure type. No superconductivity above 2 K was detected in $Zr_5Sn_3$, attributable to its low density of states at the Fermi-level and absence of strongly phonon-coupled bands around the Fermi energy given by the band structure calculations. Detailed doping studies have shown that superconductivity in $Zr_5Sb_3$ is rather robust with Hf- and Y-substitution of Zr, but suppressed by Ti-substitution; it is also suppressed by interstitial filling in $Zr_5Sb_3Z$ by Z = Sb, C, or O. In view of the large number of compounds in the $Mn_5Si_3$ structure family, more superconductors, with some of higher $T_c$, are expected. An extensive search for superconductivity in this and related compound families is underway.

Acknowledgments

The work in Houston is supported in part by U.S. Air Force Office of Scientific Research (AFOSR) contract FA9550-09-1-0656, the T. L. L. Temple Foundation, the John J. and Rebecca Moores Endowment, and the State of Texas through the Texas Center for Superconductivity at the University of Houston. Z. P. Y. and G. K. are supported by the U.S. AFOSR MURI program.